\documentclass[twocolumn,showpacs,prl,superscriptaddress]{revtex4}
\usepackage{amsfonts}
\usepackage{amssymb}
\usepackage{amsmath}
\usepackage{mathrsfs}
\usepackage{graphicx}
\usepackage{array}
\usepackage{bm}
\usepackage[dvipdfm]{hyperref}
\usepackage{mathptmx}

\begin{document}

\title{Electrically controllable surface magnetism on the surface of
topological insulator}
\author{Jia-Ji Zhu}
\affiliation{SKLSM, Institute of Semiconductors, Chinese Academy of Sciences, P. O. Box
912, Beijing 100083, China}
\author{Dao-Xin Yao}
\affiliation{School of Physics and Engineering, Sun Yat-sen University, Guangzhou 510275,
China}
\author{Shou-Cheng Zhang}
\affiliation{Department of Physics, Stanford University, CA 94305, USA}
\author{Kai Chang}
\affiliation{SKLSM, Institute of Semiconductors, Chinese Academy of Sciences, P. O. Box
912, Beijing 100083, China}

\begin{abstract}
We study theoretically the RKKY interaction between magnetic impurities on
the surface of three-dimensional topological insulators, mediated by the
helical Dirac electrons. Exact analytical expression shows that the RKKY
interaction consists of the Heisenberg-like, Ising-like and DM-like terms.
It provides us a new way to control surface magnetism electrically. The gap
opened by doped magnetic ions can lead to a short-range Bloembergen-Rowland
interaction. The competition among the Heisenberg, Ising and DM terms leads
to rich spin configurations and anomalous Hall effect on different lattices.
\end{abstract}

\maketitle

\emph{Introduction.---} All-electrical quantum manipulation of the spin
degree of freedom of electrons and/or magnetic ions is a central issue in
the fields of spintronics and quantum information processing. Spin-orbit
interactions in solids is a inherent relativistic effect and can be tuned by
an external electric field, i.e., breaking of spatial inversion symmetry.
Recent discovery demonstrate that the spin-orbit interactions can invert the
conduction and valence bands, and leads to a new state of matter named
topological insulator (TI) in a number of materials, such as a
two-dimensional (2D) HgTe quantum well \cite{BHZ,Molenkamp,Chang}, and
three-dimensional (3D)\ Bi$_{x}$Sb$_{1-x}$ \cite{Fu1,Hsieh1}, Bi$_{2}$Se$%
_{3} $ and Bi$_{2}$Te$_{3}$ \cite{HJZhang,Xia,Chen}. These TIs possess an
insulating bulk and metallic edge or surface states that are protected by
the time reversal symmetry (TRS). The energy spectrum of the surface states
shows a single massless Dirac cone at $\Gamma $ point of the Brillouin zone,
which was experimentally verified using the angle-resolved photoemission
spectroscopy (ARPES) experiments \cite{Hsieh1,Chen,Hsieh2}. Electrical
tuning of the Fermi energy in 3D TI materials is reported by changing the
backgate voltage \cite{YQLi}. Although the energy spectrum of the surface
states of 3D TI is very similar with that of graphene, but there is an
important difference between graphene and TIs. The helicity of graphene is
not defined in regard to the real spin of the electron but to the two
sublattices of graphene. Contrastingly, the helicity of the surface state of
3D TI is defined in regard to the electron spin. The helical surface state
is spin-momentum locked and certainly leads to spin relevant effects, e.g.,
spin filtering and giant magnetoresistance, the
Ruderman-Kittel-Kasuya-Yosida (RKKY) interaction \cite{Liu,Ye}.

The RKKY interaction is an indirect exchange interaction between magnetic
ions mediated by itinerant electrons. This long-range spin-spin interaction
play a crucial role in magnetic metals and diluted magnetic semiconductor.
One can expect that the mediated helical electrons on the surface of a 3D TI
must lead to some unusual properties of the RKKY interaction. Importantly,
introducing of many magnetic impurities in 3D TIs breaks the TRS and open a
gap in the Dirac spectrum of the surface states \cite{Shen}. This transition
makes the massless Dirac electrons become the massive ones and change the
spin orientation of the surface states near the Dirac point. The
characteristics of the itinerant electron states, e.g., chirality and energy
dispersion, determine the spin-spin interaction between magnetic impurities.
However, the back action of magnetic impurities also affects the electron
states, e.g., the gap opening, and consequently alters the spin-spin
interaction. Recent experiments demonstrate that a strong warping effect in
the energy spectrum of the surface states of the 3D TIs, e.g., Bi$_{2}$Se$%
_{3}$ and Bi$_{2}$Te$_{3}$ \cite{Chen,Hsieh2,Kimura}. The warping effect
becomes more significant at high Fermi energy, and certainly influences the
surface magnetism of 3D TIs.

In this Letter, we draw attention to the surface magnetism of a 3D TI,
utilizing the RKKY interaction mediated by helical massless or massive Dirac
electrons. From the analytical expression of the RKKY interaction, we
demonstrate theoretically that one can implement various quantum spin models
by changing applied gate voltage, e.g., the Dzyaloshinskii-Moriya (DM)
model, the XXZ model and the XZ model. The gap opening caused by the
magnetic impurities results in an additional Ising term in the RKKY
interaction and lead to a short-range correlation, i.e., the
Bloembergen-Rowland (BR) interaction, when the Fermi energy is located in
the gap. The warping effect behaves like an anisotropic momentum-dependent
effective magnetic field perpendicular to the surface, leading to an
crystallographic orientation-dependent RKKY interaction. The local spins can
be arranged in various lattices, e.g., triangular and square lattices formed
by the STM technique, the pinning effect, or the Coulomb interaction. The
interplay between the unique property of the RKKY interaction and the
geometry of spin lattice results in the rich spin configurations of the
ground states of spin systems, e.g., the ferromagnetic, antiferromagnetic
and spin frustration on the surface of a 3D TI.

\begin{figure}[tph]
\includegraphics[width=\columnwidth]{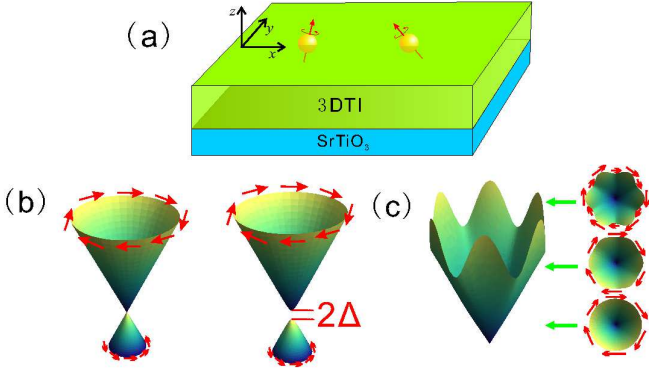}
\caption{(color online). (a) Schematic of two local spins on the surface of
TI, with electrically controllable RKKY interaction. The 3D topological thin
film is epitaxially grown on SrTiO$_{3}$ substrates \protect\cite{YQLi}. (b)
The massless and massive surface state with spin orientation. The $2\Delta $
denotes the gap caused by breaking TRS. (c) The warping effect of Bi$_{2}$X$%
_{3}$ class TI with spin orientation.}
\label{fig:fig1}
\end{figure}

\emph{Hamiltonian and RKKY interaction.---}First we consider two magnetic
impurities $\mathbf{S}_{i}$ $\left( i=1,2\right) $ located at $\mathbf{R}%
_{i} $ mediated by massless Dirac electrons on the surface of 3D TI (see
Fig. \ref{fig:fig1}(a)). The Hamiltonian of the system is $%
H=H_{0}+H_{i}^{int}$, where the massless Dirac electron Hamiltonian \cite%
{Fu2} $H_{0}=\hbar v_{F}\left( k_{x}\sigma _{y}-k_{y}\sigma _{x}\right) +%
\frac{\lambda }{2}\left( k_{+}^{3}+k_{-}^{3}\right) \sigma _{z}$ and the
\textit{s-d} interaction between the magnetic impurities and the electrons $%
H_{i}^{int}=-J\left( \overrightarrow{\sigma }\mathbf{\cdot S}\right) \delta
\left( \mathbf{r}-\mathbf{R}_{i}\right) $. $\mathbf{k}$ is in-plane momentum
of electron, $\overrightarrow{\sigma }=(\sigma _{x},\sigma _{y},\sigma _{z})$
is the Pauli matrix denoting real spin of electron. $v_{F}$ is the Fermi
velocity of the surface states, which is given by $6.2\times 10^{5}$ms$^{-1}$
for Bi$_{2}$Se$_{3}$ \cite{HJZhang} and $3.9\times 10^{5}\mathrm{ms}^{-1}$
for Bi$_{2}$Te$_{3}$ \cite{Fu2}. $\widehat{z}$ is a unit vector along the
normal direction of the surface. $\lambda $ is the magnitude of the
hexagonal warping term. $J$ denotes the strength of the \textit{s-d}
exchange interaction, which is around $2\backsim 13$ eV$\overset{\circ }{%
\mathrm{A}}^{2}$\cite{Liu}.

We first neglect the warping effect, the Green's function in real space can
be obtained as $G_{\varepsilon }\left( \pm \mathbf{R}\right) =-\frac{%
\varepsilon }{4\hbar ^{2}v_{F}^{2}}\left[ iH_{0}^{\left( 1\right) }\left(
\frac{R\varepsilon }{\hbar v_{F}}\right) \mp \widehat{z}\cdot \left(
\overrightarrow{\sigma }\mathbf{\times n}_{R}\right) H_{1}^{\left( 1\right)
}\left( \frac{R\varepsilon }{\hbar v_{F}}\right) \right] $, where $H_{\nu
}^{\left( 1\right) }\left( x\right) $ is the $\nu $-order Hankel function of
the first kind, and $\mathbf{R}\equiv R\mathbf{n}_{R}$. In the loop
approximation we find the RKKY interaction between two magnetic impurities
in the form of $H_{1,2}^{RKKY}=-\frac{2}{\pi }$\textrm{Im}$\int_{-\infty
}^{\varepsilon _{F}}\mathrm{d}\varepsilon $\textrm{Tr}$[H_{1}^{int}G_{%
\varepsilon }\left( \mathbf{R};\varepsilon +i0^{+}\right)
H_{2}^{int}G_{\varepsilon }\left( -\mathbf{R};\varepsilon +i0^{+}\right) ]$,
where $\varepsilon _{F}$ is the Fermi energy, and $\mathrm{Tr}$ means a
partial trace over the spin degree of freedom of itinerant Dirac electrons.
Then the RKKY interaction can be written as%
\begin{eqnarray}
H_{1,2}^{RKKY} &=&F_{1}\left( R,\varepsilon _{F}\right) \mathbf{S}_{1}%
\mathbf{\cdot S}_{2}+F_{2}\left( R,\varepsilon _{F}\right) \left( \mathbf{S}%
_{1}\mathbf{\times S}_{2}\right) _{y}  \notag \\
&&+F_{3}\left( R,\varepsilon _{F}\right) S_{1}^{y}S_{2}^{y},  \label{gapless}
\end{eqnarray}%
where the range functions are

\begin{eqnarray*}
F_{1}\left( R,\varepsilon _{F}\right) &=&\frac{J^{2}}{4\pi ^{\frac{3}{2}%
}\hbar v_{F}R^{3}}{\Huge [}G_{2,4}^{2,1}\left( \frac{R^{2}\varepsilon
_{F}^{2}}{\hbar ^{2}v_{F}^{2}}|%
\begin{array}{c}
1,2 \\
\frac{3}{2},\frac{5}{2},0,\frac{1}{2}%
\end{array}%
\right) \\
&&-G_{2,4}^{2,1}\left( \frac{R^{2}\varepsilon _{F}^{2}}{\hbar ^{2}v_{F}^{2}}|%
\begin{array}{c}
1,2 \\
\frac{3}{2},\frac{3}{2},0,\frac{3}{2}%
\end{array}%
\right) +\frac{\sqrt{\pi }}{2}{\Huge ]},
\end{eqnarray*}

\begin{eqnarray*}
F_{2}\left( R,\varepsilon _{F}\right) &=&\frac{J^{2}}{4\pi ^{\frac{3}{2}%
}\hbar v_{F}R^{3}}{\Huge [}G_{2,4}^{2,1}\left( \frac{R^{2}\varepsilon
_{F}^{2}}{\hbar ^{2}v_{F}^{2}}|%
\begin{array}{c}
2,\frac{3}{2} \\
2,2,0,1%
\end{array}%
\right) \\
&&-G_{1,3}^{2,0}\left( \frac{R^{2}\varepsilon _{F}^{2}}{\hbar ^{2}v_{F}^{2}}|%
\begin{array}{c}
\frac{3}{2} \\
1,2,0%
\end{array}%
\right) {\Huge ]},
\end{eqnarray*}

\begin{equation*}
F_{3}\left( R,\varepsilon _{F}\right) =\frac{J^{2}}{2\pi ^{\frac{3}{2}}\hbar
v_{F}R^{3}}{\Huge [}G_{2,4}^{2,1}\left( \frac{R^{2}\varepsilon _{F}^{2}}{%
\hbar ^{2}v_{F}^{2}}|%
\begin{array}{c}
1,2 \\
\frac{3}{2},\frac{3}{2},0,\frac{3}{2}%
\end{array}%
\right) -\frac{3\sqrt{\pi }}{8}{\Huge ]}.
\end{equation*}%
Here $G_{p,q}^{m,n}$ is the Meijer's G-function \cite{TISP}. The RKKY
interaction consists of three terms: the Heisenberg-like term, the DM-like
term and the Ising-like term, displaying different range functions. The
competition among these three terms can implement various spin models.

\begin{figure}[tph]
\includegraphics[width=\columnwidth]{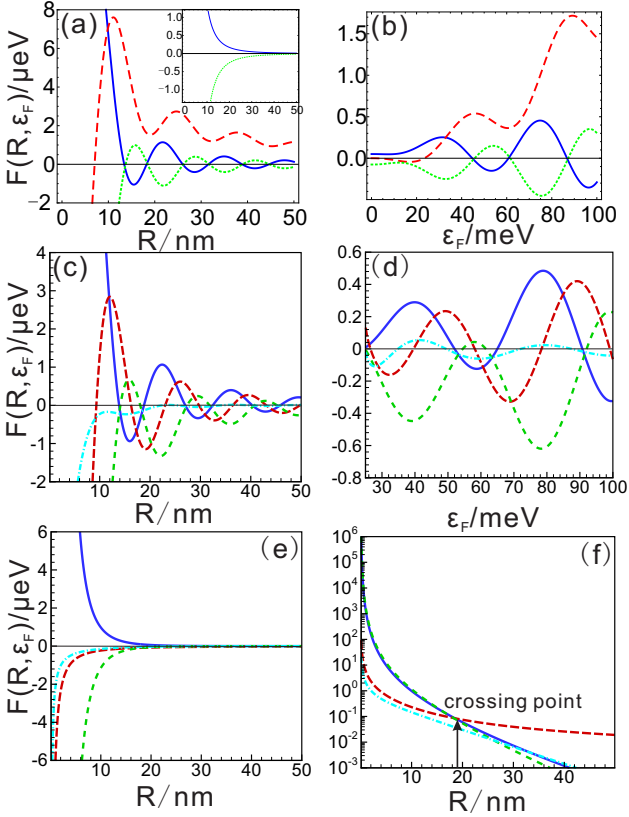}
\caption{(color online). The range function of RKKY interaction as function
of the distance between localized spins $R$ (Fig.2(a,c)) or the Fermi energy
$\protect\varepsilon _{F}$ (Fig.2(b,d)). The inset shows the intrinsic case (%
$\protect\varepsilon _{F}=0$). Fig. 2(a),(b) depict the RKKY interaction
with massless Dirac electrons, and Fig. 2(c),(d) the RKKY interaction with
massive electrons ($\Delta =25$ meV)\protect\cite{Shen}. Fig. 2(e) is the
range functions of BR interaction, and Fig. 2(f) the absolute values of BR
range functions with logarithmic coordinates. The blue (solid) lines are the
range function $F_{1}(R,\protect\varepsilon _{F})$, the red (long dashed)
lines are the $F_{2}(R,\protect\varepsilon _{F})$, the green (short dashed)
ones are $F_{3}(R,\protect\varepsilon _{F})$, and the cyan (dash dotted)
ones are $F_{4}(R,\protect\varepsilon _{F})$. We choose $\protect\varepsilon %
_{F}=100\text{ meV}$ for Fig. 2(a,c), $R=30\text{ nm}$ for Fig. 2(b,d) and $%
\protect\varepsilon _{F}=-18$ meV for Fig. 2(e,f). The zero point of energy
is at the midpoint of the energy gap caused by breaking TRS in Fig.
2(c)-(f). }
\label{fig:RKKY}
\end{figure}

All three range functions show damped oscillation as the distance $R$
increases, this behavior indicates that this RKKY interaction is a
long-range correlation between two spins (see Fig. \ref{fig:RKKY}(a)).
Utilizing the asymptotic form of Hankel functions, the long-range asymptotic
behavior of the RKKY\ interaction, i.e., $\varepsilon _{F}R\gg 1$, is
\begin{eqnarray}
H_{1,2}^{RKKY\left( asym\right) } &\approx &\frac{J^{2}\varepsilon _{F}}{%
2\pi ^{2}\hbar ^{2}v_{F}^{2}R^{2}}{\Large [}\sin \left( \frac{2R\varepsilon
_{F}}{\hbar v_{F}}\right) \left( \mathbf{S}_{1}\mathbf{\cdot S}%
_{2}-S_{1}^{y}S_{2}^{y}\right)  \notag \\
&&-\cos \left( \frac{2R\varepsilon _{F}}{\hbar v_{F}}\right) \left( \mathbf{S%
}_{1}\mathbf{\times S}_{2}\right) _{y}{\Large ].}  \label{Asym}
\end{eqnarray}%
From Eq. (\ref{Asym}), the oscillatory behavior of RKKY interaction in a
\textit{n-}type TI ($\varepsilon _{F}>0$) can be clearly seen and dominates
at large $\varepsilon _{F}R$. The long-range behaviors of the Heisenberg
term and Ising term show a spatial dependence as $1/R^{2}$, which is the
same as in a conventional 2D electron gas, but in contrast with that in
graphene where the spatial dependence is $1/R^{3}$. Interestingly, one can
see that the range functions for the Heisenberg and Ising terms have almost
the same magnitude but opposite sign. It means that the Ising term always
cancel the \textit{z}-component of the Heisenberg term. By properly
adjusting the distance $R$ using the STM technique and/or the Fermi energy $%
\varepsilon _{F}$ {(}see Fig. \ref{fig:RKKY}(b){)}, one can even diminish
the Heisenberg and Ising terms and obtain a RKKY interaction containing the
DM\ interaction alone. The RKKY interaction becomes $H_{1,2}^{RKKY}\left(
R,\varepsilon _{F}^{c}\right) \simeq F_{2}\left( R,\varepsilon
_{F}^{c}\right) \left( \mathbf{S}_{1}\mathbf{\times S}_{2}\right) _{y}$,
i.e., a pure DM-like spin model. The long-range asymptotic form of the DM\
interaction also decreases as $1/R^{2}$, but with $\pm \pi /2$ phase shift
compared to other two terms. By setting $\varepsilon _{F}=0$, i.e., the
intrinsic 3D TI case, the DM term vanishes, the RKKY interaction can be
simply written as $H_{1,2}^{RKKY}=\frac{J^{2}}{16\pi \hbar v_{F}R^{3}}\left(
2S_{1}^{x}S_{2}^{x}+2S_{1}^{z}S_{2}^{z}-S_{1}^{y}S_{2}^{y}\right) $, a
XXZ-like spin model. Notice that the range function monotonically decreasing
as $1/R^{3}$ (see the inset in Fig. \ref{fig:RKKY}(a)) in this case, the
same as that in graphene.

\emph{The gap opening and warping effects.---}Next we study the gap opening
effect caused by magnetic impurities in the TIs. The surface states are
described by a massive Dirac Hamiltonian $H_{0}^{\left( gap\right) }=\hbar
v_{F}\left( k_{x}\sigma _{y}-k_{y}\sigma _{x}\right) +\Delta \sigma _{z}$,
where $2\Delta $ denotes the gap caused by breaking TRS (see Fig. \ref%
{fig:fig1}(b)). This Hamiltonian means that the magnetic impurities behave
as a Zeeman term which lifts the degeneracy of Dirac point. Similarly, we
get the RKKY interaction including the gap opening effect,
\begin{eqnarray}
H_{1,2}^{RKKY\left( gap\right) } &=&F_{1}^{g}\left( R,\varepsilon
_{F}\right) \mathbf{S}_{1}\mathbf{\cdot S}_{2}+F_{2}^{g}\left( R,\varepsilon
_{F}\right) \left( \mathbf{S}_{1}\mathbf{\times S}_{2}\right) _{y}  \notag \\
&&+F_{3}^{g}\left( R,\varepsilon _{F}\right)
S_{1}^{y}S_{2}^{y}+F_{4}^{g}\left( R,\varepsilon _{F}\right)
S_{1}^{z}S_{2}^{z}.  \label{gap}
\end{eqnarray}%
There are two dominant differences between the RKKY interaction for the
gapped (Eq.(\ref{gap})) and gapless (Eq.(\ref{gapless})) cases when the
Fermi energy is located in the conduction band ($\varepsilon _{F}>\Delta $)
(see Fig. \ref{fig:RKKY}(c,d)). The first is that the range function of the
DM term can be negative for the gapped case. We can still implement the pure
DM model by properly choosing a specific Fermi energy $\varepsilon _{F}^{c}$
and/or the spacing of neighboring spins. The second difference is that the
gap opening induces an additional out-of-plane Ising term along the $z$\
direction with a relatively weak correlation between the local spins. This
is because the gap opening effect looks like a perpendicular magnetic field
leading to a Zeeman-like splitting $\Delta \sigma _{z}$.

When we tune the Fermi energy into the gap, the RKKY interaction can be a
short-range interaction, i.e., the BR interaction \cite{BR}. Figs \ref%
{fig:RKKY}(e) and \ref{fig:RKKY}(f) show that the BR interaction mediated by
massive Dirac electrons also contains all the four terms. This is because
the spin orientation on the surface of constant is not influenced by the gap
opening away from the vicinity of Dirac point. However, the range functions
are totally different. From Fig. \ref{fig:RKKY}(e) we can see the Heisenberg
term is always antiferromagnetic, while the two Ising terms always give
ferromagnetic correlation. Fig. \ref{fig:RKKY}(f) shows the exponentially
spatial decay behavior of the BR interaction. The Heisenberg term $%
F_{1}^{g}\left( R,\varepsilon _{F}\right) $ and Ising term $F_{3}^{g}\left(
R,\varepsilon _{F}\right) $ are dominant with almost the same amplitude
under a crossing point of spacing $R\simeq 19$ nm, which means one can
realize the XZ model $H_{1,2}^{BR}\simeq F_{1}^{g}\left( R,\varepsilon
_{F}\right) \left( S_{1}^{x}S_{2}^{x}+S_{1}^{z}S_{2}^{z}\right) $ on the
lattice of the surface of TI. Above this crossing point DM term play the
leading role, however, with rather small values.

\begin{table*}[tbph]
\caption{The crystallographic orientation-dependent RKKY interaction caused
by the warping effect. The range functions depend on the angle $\protect%
\theta $ which is defined in the direction respect to the [110]
crystallographic direction, and are different from the range functions without warping effect.}
\label{tab:warping}%
\begin{ruledtabular}
\begin{tabular}{cl}
Crystallographic directions&\multicolumn{1}{c}{$H_{1,2}^{RKKY}$}\\

\hline

$[110]$&$F_{1}^{w}\left( R,\theta,\varepsilon _{F}\right) \mathbf{S}_{1}\mathbf{%
\cdot S}_{2}+F_{2}^{w}\left( R,\theta,\varepsilon _{F}\right)
\left( \mathbf{S}_{1}\mathbf{\times S}_{2}\right) _{y}+F_{3}^{w}\left(R,\theta,\varepsilon _{F}\right) S_{1}^{y}S_{2}^{y}+F_{4}^{w}\left(
R,\theta,\varepsilon _{F}\right) S_{1}^{z}S_{2}^{z}$\\

$[\overline{1}10]$&$F_{1}^{w}\left( R,\theta,\varepsilon
_{F}\right) \mathbf{S}_{1}\mathbf{\cdot S}_{2}+F_{2}^{w}\left( R,\theta,\varepsilon _{F}\right) \left( \mathbf{S}_{1}\mathbf{\times S}%
_{2}\right) _{y}+F_{3}^{w}\left( R,\theta,\varepsilon _{F}\right)
S_{1}^{y}S_{2}^{y}$\\

$[010]$&$F_{1}^{w}\left( R,\theta,\varepsilon _{F}\right) \mathbf{S}_{1}\mathbf{\cdot S}_{2}+%
F_{2}^{w}\left( R,\theta,\varepsilon _{F}\right) \left( \mathbf{S}_{1}%
\mathbf{\times S}_{2}\right) _{z}+F_{3}^{w}\left( R,\theta,\varepsilon _{F}\right) S_{1}^{y}S_{2}^{y}
+F_{4}^{w}\left( R,\theta,\varepsilon _{F}\right) S_{1}^{z}S_{2}^{z}
+F_{5}^{w}\left( R,\theta,\varepsilon _{F}\right) S_{1}^{x}S_{2}^{x}$\\

& $+F_{6}^{w}\left( R,\theta,\varepsilon _{F}\right) \left(
S_{1}^{x}S_{2}^{y}+S_{1}^{y}S_{2}^{x}\right) $\\
\end{tabular}
\end{ruledtabular}
\end{table*}

With increasing the Fermi energy, the warping effect of the energy
dispersion of the surface states becomes more significant, resulting in a
hexagonal constant energy surface {(}see Fig. \ref{fig:fig1}(c){)}. The
warping effect behaves like a perpendicular anisotropic effective magnetic
field that breaks the rotational symmetry, which is different from the gap
opening effect. This anisotropy must lead to an anisotropic, i.e., a
crystallographic orientation-dependent RKKY interaction. We consider two
local spins arranged along different crystallographic directions (see Table %
\ref{tab:warping}). These formula indicate that the anisotropic effective
magnetic field results in more complicated spin models.

\begin{figure}[tph]
\includegraphics[width=\columnwidth]{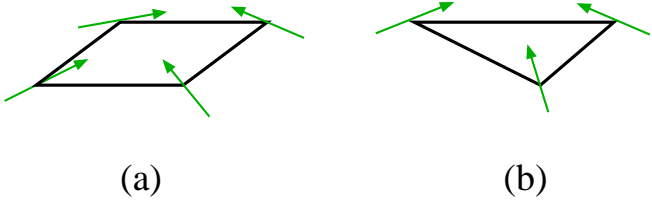}
\caption{(color online). Schematic chiral spin configurations of spin
systems in the square and triangular lattices with the DM interaction and
Heisenberg interaction.}
\label{fig:fig3}
\end{figure}

\emph{Spin configurations and anomalous Hall effect.---}Now we turn to
discuss the possible spin configurations of the ground states of the 2D spin
systems in the square and triangulated lattices on the surface of 3D TI. The
combination of the Heisenberg interaction, DM interaction, and Ising
interactions makes it a rich platform to study all kinds of spin
configurations. First we consider a pure DM model which can be implemented
by tuning the Fermi energy or the spacing of neighboring spins. The DM
interaction twist the spin orientation of the neighboring spins and leads to
a non-coplanar spin phase or chiral spin phase. For example, the DM
interaction cant the spins out of the plane and form an umbrella-like spin
structure with a net ferromagnetic moment (see Fig. \ref{fig:fig3}). When
chiral electrons hop on the non-coplanar spin sites, the anomalous Hall
effect can be realized since they obtain complex phase factor (Berry phase)
which works as an internal magnetic field if it is not canceled. Thus the
presence of spin chirality and net ferromagnetic moment can yield an
anomalous Hall effect on the surface of the 3D TI \cite{Wen,Nagaosa2}. In a
square spin lattice, the pure DM interaction can lead to a perpendicular
nearest neighbor spin ordering. Together with the Heisenberg interaction,
the spin ordering can be non-coplanar in the square lattice shown in Fig. %
\ref{fig:fig3} (a) . For a triangulated spin lattice, the DM term and
ferromagnetic Ising terms can induce spin frustration and have spins to
point out of the plane to form a chiral spin state, illustrated in Fig. \ref%
{fig:fig3} (b). If we define the spin chirality $\mathbf{S}_{1}\cdot (%
\mathbf{S}_{2}\times \mathbf{S}_{3})$ to describe the ferromagnetic moment
for each triangle, the anomalous Hall effect can happen since the Berry
phase is proportional to the spin chirality and may not be canceled \cite%
{Nagaosa2}. In the Kagome lattice antiferromagnet, thermodynamic and neutron
scattering measurements have revealed the chiral spin configuration and net
ferromagnetism induced by the DM interaction \cite{kagome}. The anomalous
Hall effect can be measured by the transverse Hall conductivity on the
surface of 3D TI.

When the antiferromagnetic Heisenberg interaction dominates, the spin
configuration is generally collinear or coplanar, which has no spin
chirality. The anomalous Hall effect is not expected. The square spin
lattice has long-range collinear antiferromagnetic ground state. For a
triangulated spin lattice, the long-range coplanar state may develop, like
the well known $120^o$ long range order. Since the existence of spin
frustration, the ground state of triangulated spin lattices can be highly
degenerate and show no long range order. The introduction of DM interaction
and ferromagnetic Ising interactions will remove the degeneracy partly or
completely. By finely changing the parameters, the spin system can even
transit into a spin-glass phase. This means one can construct an artificial
spin frustration system utilizing STM technique on the surface of the 3D TI.

Neutron scattering will be a useful tool to detect the surface and bulk spin
structure of 3D TI. To enhance the neutron scattering signal from the
surface magnetism, it is useful to use large angle neutron scattering
technique and pile up many layers of 3D TI with parallel surface.

In summary, we propose a new scheme to manipulate the carrier mediated
spin-spin interaction on the surface of the 3D TI electrically, i.e., RKKY
interaction and BR interaction. The helical surface state leads to the
twisted DM interaction and an in-plane Ising interaction. The gap opening
and warping effects introduce isotropic and anisotropic Ising-like terms at
low and high Fermi energies, respectively. This spin-spin interaction can be
used to realize different spin models by adjusting Fermi energy, such as DM
model, XXZ model and XZ model. These realizations would not be destroyed by
the gap opening caused by the breaking TRS and the warping effect. The
surface magnetism of 3D TIs provide us a model platform to study spin
configurations and dynamics of various spin models, and pave the way for
explore new fundamental physics and new type spintronic devices.

\begin{acknowledgments}
This work was supported by the NSFC Grant Nos. 60525405 and 10874175, and
the Knowledge Innovation Project of CAS. SCZ is supported by the NSF under
grant numbers DMR-0904264.
\end{acknowledgments}

\end{document}